\documentclass[a4paper]{jpconf}

\usepackage{amsmath,amssymb,amsthm,amsfonts}
\usepackage{graphicx} 
\usepackage{graphicx}        

\newtheorem{theorem}{Theorem}

\begin{document}

\title{Operator-algebraic construction of the deformed Sine-Gordon model}
\author{Daniela Cadamuro}
\address{Universit\"at Leipzig, Institut f\"ur Theoretische Physik, Br\"uderstra\ss e 16, 04103 Leipzig, Germany}
\ead{daniela.cadamuro@itp.uni-leipzig.de}
%
%

\begin{abstract}
We consider the construction of integrable quantum field theories in the operator-algebraic approach,  which  is  based  on  quantum  fields  localized  in  infinitely  extended  wedge  regions. 
This  approach  has  been  successful  for the construction of a class of models with scalar $S$-matrices and without bound states.  In extension of these results, we apply similar methods to $S$-matrices with poles in the physical strip (``bound states'').  Specifically, we consider a deformed version of the Sine-Gordon model, containing only breathers.  We exhibit wedge-local fields in this model, which differ from those in non-bound state models by an additive term, the so called ``bound state operator''.
\end{abstract}

\section{Introduction}
\label{sec:1}

The construction of interacting quantum field theories is a difficult problem in Mathematical Physics. A usual approach is based on the perturbative expansion of the $S$-matrix constructed from a Lagrangian, but the appearance of infrared and ultraviolet divergences in the expansion, as well as the problem of convergence of the perturbation series, make expressions mathematically ill-defined. 

There is however a class of quantum field theories that can be constructed with techniques which avoid perturbative expansions: the quantum integrable models. These are $1+1$-dimensional field theories where two-particle scattering processes characterize the theory completely. They are constructed as an inverse scattering problem: Given a function $S$ as a mathematical input, one constructs the corresponding QFT having this two-particle scattering function. The interaction of these models is simple enough for certain methods familiar from free field theory to be carried over to the interacting situation.

Despite being restricted to low dimensions, there is physical interest in studying these models. In fact, they share interesting common features with interacting theories in higher dimensions. For example, the nonlinear $O(N)$ sigma models \cite{BabujianFoersterKarowski:2013} are linked to experimentally realizable situations in condensed matter systems. They can also be regarded as simplified analogues of four-dimensional nonabelian gauge theories, inasmuch as they share crucial features with them, including renormalizability, asymptotic freedom, and the existence of instanton solutions.
%
%
In this sense, integrable systems provide a ``landscape'' of possible interactions, where one may hope to test techniques that can be adapted to, or inspire methods for, systems in higher dimensions, and where one may obtain insight into the structure of local QFTs with a less simple $S$-matrix.



Here, we will focus on a particular model of the integrable class, the Sine-Gordon model, which has been treated with several methods before. In particular, the integrability of the massless Sine-Gordon model and its $S$-matrix have been conjectured in \cite{ZamolodchikovZamolodchikov:1979}. The corresponding inverse scattering problem was studied
in the form factor programme \cite{BabujianFringKarowskiZapletal:1993,Babujian2002} where certain matrix elements of the pointlike local fields (``form factors'') have been computed; the existence of a Wightman field remains unclear though due to convergence issues in the infinite expansion of its $n$-point functions in terms of form factors \cite{BabujianKarowski:2004}. In a rigorous framework (pAQFT) it was shown \cite{BahRej16,Bahns2017} that in the massless Sine-Gordon model in the ultraviolet-finite regime (hence, without need of renormalization), the expectation value of the Epstein-Glaser $S$-matrix given as formal power series converges in a Hadamard state. The authors also constructed the interacting currents and the interacting field as formal power series in the coupling constant, and showed convergence in expectation
value in that state. Yet, the factorization of the $S$-matrix has not been proven there. We should also mention the conjectured equivalence of the massless Sine-Gordon model with the Thirring model (Coleman's equivalence), which has been proven by Benfatto, Falco and Mastropietro \cite{BFM09} in the case of finite volume interaction of the massless Sine-Gordon model and of the Thirring model with a finite volume mass term. A construction of the Thirring model by functional integral methods can be found in \cite{BFM07}. 

In this paper, we report on progress towards a construction of a deformed version of the Sine-Gordon model, which contains only ``breather'' particles (Sec.~\ref{sec:sg}). We construct quantum fields associated with this model which are---in the sense of their commutation relations---localized in spacelike wedges (Sec.~\ref{sec:wedge}), as proposed by Schroer \cite{SchroerWiesbrock:2000-1}. This extends our previous results for scalar models with bound states \cite{CT2015}. The construction of strictly local observables, which is ongoing work, would then use methods from the theory of $C^\ast$-algebras and the abstract framework of Tomita-Takesaki modular theory, as already known for the Sinh-Gordon model \cite{Lechner:2008}. 

This proceeding is based on joint work with Y.~Tanimoto \cite{CTsine}.

\section{Properties of the $S$-matrix}\label{sec:2}

Quantum integrable models have the property that the $S$-matrix is factorizable into two-particle scattering matrices. This two-particle scattering matrix, a meromorphic matrix-valued function (in the following only referred to as $S$), needs to fulfill a number of axioms, including \emph{unitarity},  \emph{crossing symmetry}, the \emph{Bootstrap equation} and, in the case of several particle species, the \emph{Yang-Baxter equation}.
In some cases, only matrix elements of the form $S_{\alpha \beta}^{\beta \alpha}$ are nonzero; in physical terms, two particles of type $\alpha, \beta$ can scatter with the exchange of a unitary factor, but the type of particles stays the same in scattering. In this case, the $S$-matrix is said to be ``diagonal''. There are also ``non-diagonal'' $S$-matrices where the type of particle can change due to scattering. 

We are interested in the class of integrable models where bound states of particles are present. Mathematically, these bound states correspond to poles of $S$ in the \emph{physical strip} (the region $0\leq \operatorname{Im} \zeta \leq \pi$ in complex rapidity space). We denote with $\theta_1, \theta_2$ the rapidities of the particles $\alpha, \beta$, which parametrize their momenta. These two particles can fuse into a third particle of type $\gamma$ in the following sense:
\begin{equation}
\pmb{p}_\alpha (\theta +i\theta_{(\alpha \beta)}) + \pmb{p}_\beta (\theta - i \theta_{(\beta \alpha)}) = \pmb{p}_\gamma (\theta),
\end{equation}
i.e., the momenta of two virtual particles add to the momentum of a third real particle (the bound particle) which lies on the mass shell. The numbers $\theta_{(\alpha \beta)}, \theta_{(\beta \alpha)}$ are determined by this equation, once the masses of the particles are fixed. The bound state formed would correspond to a pair of simple poles of the component $S_{\alpha \beta}^{\beta \alpha}(\zeta)$ in the physical strip: the s-channel pole, located at $\zeta = i\theta_{\alpha \beta} := i\theta_{(\alpha \beta)} +i\theta_{(\beta \alpha)}$, and the t-channel pole at $\zeta = i\pi - i\theta_{\alpha \beta}$, which arises due to crossing symmetry. The numbers $\theta_{\alpha \beta}$ are called \emph{fusion angles}. The possible fusion processes, which we denote by the symbol $(\alpha \beta) \to \gamma$, are characteristic of each model, and listed in its fusion table.

\section{The deformed Sine-Gordon model}\label{sec:sg}

The particle spectrum of the Sine-Gordon model contains solitons ($s$), anti-solitons ($\bar s$) and several breathers ($b_k$, $1 \leq k \leq K$), which are neutral particles  (i.e., $\overline{b_k} = b_k$).
The number $K$ of breathers is the largest integer such that $K \nu <2$, where $0< \nu<1$ is a coupling constant. 
The fusion processes between these particles and corresponding fusion angles are shown in Table~\ref{tab:usual}. The corresponding two-particle scattering matrix  $S_{\text{SG}}$ is of non-diagonal type or, more precisely, block-diagonal: On the soliton-antisoliton subspace, it has the form
\begin{equation}
S_{\text{SG,soliton}}(\zeta) =\begin{pmatrix}
a(\zeta) & 0 & 0 & 0 \\
0 & t(\zeta) & r(\zeta)& 0 \\
0 & r(\zeta) & t(\zeta) & 0 \\
0 & 0 & 0 & a(\zeta)
\end{pmatrix},
\end{equation}
where 
\begin{equation}
a(\zeta) := S^{ss}_{ss}(\zeta) = S^{\bar s  \bar s}_{\bar s \bar s}(\zeta), \quad  t(\zeta) := S^{\bar s s}_{s \bar s}(\zeta) = S^{s \bar s}_{\bar s s}(\zeta), \quad r(\zeta):= S^{s \bar s}_{s \bar s}(\zeta) = S^{\bar s s}_{\bar s s}(\zeta);
\end{equation}
the expressions $a(\zeta), t(\zeta), r(\zeta)$, depending on $\nu$, are explicitly given in \cite[Section~3.3]{Que99}. On the breather subspace, $S_{\text{SG}}$ is diagonal \cite{Que99}. 

\begin{table}
\begin{center}
\begin{tabular}{|c|c|c|}
 \hline
 processes & rapidities of particles & fusion angles \\
 \hline
 $(s \bar s) \longrightarrow b_k $ & $\theta_{(s s)}^{b_k} =\frac{\pi}{2}(1 - \nu k)$ & $\theta_{s \bar s}^{b_k} = \pi(1- \nu k)$ \\
 \hline
 $(s b_k) \longrightarrow s$ &
$\theta_{(s b_k)}^{s} =\pi \nu k, \theta_{(b_k s)}^{s} = \frac{\pi}{2}(1 - \nu k)$ & $\theta_{s b_k}^{s} = \frac{\pi}{2} \left( 1+\nu k \right)$ \\
 \hline
 $(\bar s b_k) \longrightarrow \bar s$ &
$\theta_{(\bar s b_k)}^{\bar s} =\pi \nu k, \theta_{(\bar s b_k)}^{b_k} = \frac{\pi}{2}(1 - \nu k)$ & $\theta_{\bar s b_k}^{\bar s} = \frac{\pi}{2} \left( 1+\nu k \right)$ \\
 \hline
$(b_k b_\ell) \longrightarrow b_{k + \ell}$ &
$\theta_{(b_k b_\ell)}^{b_{k + \ell}} =\frac{\pi \nu}{2}\ell, \theta_{(b_\ell b_k)}^{b_{k+\ell}} = \frac{\pi \nu}{2}k$ & $\theta_{b_k b_\ell}^{b_{k + \ell}} = \frac{\pi \nu}{2} \left( k + \ell \right)$ \\
 \hline
$(b_{k + \ell} b_k) \longrightarrow b_\ell$ &
$\theta_{(b_{k+\ell} b_k)}^{b_\ell} =\frac{\pi \nu}{2}k, \theta_{(b_k b_{k+ \ell})}^{b_\ell} = \pi \left( 1- \frac{\nu}{2}(k + \ell) \right)$ & $\theta_{b_{k+ \ell} b_k}^{b_\ell} = \pi  \left( 1 -\frac{\nu}{2} \ell \right)$ \\
 \hline
\end{tabular}
\caption{Fusion processes in the (usual) Sine-Gordon model} \label{tab:usual}
\end{center}
\end{table}

\begin{table}
\begin{center}
\begin{tabular}{|c|c|c|}
 \hline
 processes & rapidities of particles & fusion angles \\
 \hline
 $(b_1 b_1) \longrightarrow b_2 $ & $\theta_{(b_1 b_1)}^{b_2} = \frac{\pi \nu }2$ & $\theta_{b_1 b_1}^{b_2} = \pi \nu$ \\
 \hline
 $(b_2 b_1) \longrightarrow b_1, (b_1 b_2) \longrightarrow b_1$ &
$\theta_{(b_1 b_2)}^{b_1} =\pi(1 - \nu), \theta_{(b_2 b_1)}^{b_1} = \frac{\pi \nu}2$ & $\theta_{b_2 b_1}^{b_1} = \theta_{b_1 b_2}^{b_1} = \pi \left( 1-\frac{\nu}{2} \right)$ \\
 \hline
 $(b_2 b_2)$ not a fusion & &   \\
 \hline
\end{tabular}
\caption{Fusion processes in the deformed Sine-Gordon model} \label{tab:fusion}
\end{center}
\end{table}

In this paper, we will not deal with the massless Sine-Gordon model itself but with a new model with the same fusion structure, which is derived as a deformation of the ``breather-breather'' $S$-matrix of the massless Sine-Gordon model. To that end, we restrict the value of the coupling constant to $\frac{2}{3}< \nu < \frac{4}{5}$, so that there are exactly two breathers, $b_1, b_2$. 
We now consider the two-particle scattering function
\begin{equation}\label{eq:sgdeformed}
S^{b_\ell b_k}_{b_k b_\ell}(\zeta) = S_{\text{SG}}\phantom{}^{b_\ell b_k}_{b_k b_\ell}(\zeta) S_{\text{CDD}}\phantom{}^{b_\ell b_k}_{b_k b_\ell}(\zeta) , \quad \ell, k =1,2, 
\end{equation}
where the ``CDD factors'' $S_{\text{CDD}}$ are explicitly given in \cite[Sec.~2.2]{CTsine}; they are so arranged that the required properties of the function $S$, such as unitarity and crossing symmetry, are retained. 
Thus we obtain an $S$-matrix of diagonal type with particle spectrum $b_1,b_2$; we do not consider (anti-)solitons.

Each of the components $S^{b_1 b_1}_{b_1 b_1}(\zeta)$, $S^{b_1 b_2}_{b_2 b_1}(\zeta)$ and $S^{b_2 b_1}_{b_1 b_2}(\zeta)$ has only two simple poles in the physical strip (an s-channel pole as shown in Table~\ref{tab:fusion}, and a t-channel pole at $i\pi$ minus that value), and no higher-order poles in that strip. 
The corresponding fusion processes are shown in Table~\ref{tab:fusion}; note that the fusion table for $b_1,b_2$ is closed under fusions. We have \emph{maximal analyticity} for the particle $b_1$, i.e., all above-mentioned poles of $S$  correspond to fusion processes.  The poles in the component $S^{b_2 b_2}_{b_2 b_2}(\zeta)$ will not matter in our construction of the model in Sec.~\ref{sec:wedge}.

\begin{table}
\begin{center}
\begin{tabular}{|c|c|l|}
 \hline
 Range of $\nu$ & Residue of pole of $S_{b_1 b_1}^{b_1 b_1}$ & Comment\\
 \hline
 $4/5<\nu < 1$ & $-i\mathbb{R}_+$ & No adjusting CDD factors found. \\
 \hline
 $ 2/3<\nu < 4/5$ & $-i\mathbb{R}_+$ & Adjusting CDD factors found. \\
 \hline
 $ 1/2 <\nu < 2/3$ &
$-i\mathbb{R}_+$ & There are three breathers if one requires the \\
& & maximal analyticity within breathers. \\
& & No adjusting CDD factors found.\\
 \hline
 $0<\nu < 1/2$ &
$i\mathbb{R}_+$ & There is a breather $b_k$ for which \\
  & & $\displaystyle{\operatorname{res}_{\zeta = i\theta_{b_1 b_k}^{b_{k+1}}}S_{b_1b_k}^{b_kb_1}(\zeta) \in -i\mathbb{R}_+}$. \\
 \hline
\end{tabular}

\caption{Sign of the residue of $S_{b_1 b_1}^{b_1 b_1}$ for different values of $\nu$}\label{tab:cdd}

\end{center}
\end{table}
 
 The reason for including the CDD factor is that, for our proof of wedge locality of quantum fields (see Sec.~\ref{sec:wedge}), the residue of the poles in the physical strip of those $S$-components involving $b_1$ must have a certain sign, namely we need,
 \begin{equation}\label{eq:respositive}
  \operatorname{res}_{\zeta = i \theta^{b_\ell}_{b_1 b_k}}S^{b_k b_1}_{b_1 b_k}(\zeta) \in i\mathbb{R}_+ 
 \end{equation}
(\emph{positivity of residue} for $b_1$). For different values of the coupling constant $0<\nu<1$, it may or may not be possible to find such CDD factors that adjust the signs of the residues.
 We summarize the situation in Table~\ref{tab:cdd}. Due to the special properties of particle $b_1$ with respect to poles and residues of $S$, we call it an ``elementary particle" (the precise definition can be found in \cite[Section~2.1]{CT17}); this again will play an important role in Sec.~\ref{sec:wedge}.

\section{Wedge-local fields}\label{sec:wedge}

We now aim to construct a quantum field theory that is associated with the the two-particle scattering matrix \eqref{eq:sgdeformed}. We wish to bypass problems that arise when directly constructing point-local quantum fields, and hence focus on objects with weaker localization properties, so-called wedge local fields \cite{SchroerWiesbrock:2000-1}. 

More specifically: Let $W_L$ be the left and $W_R$ the right wedge in $1+1$-dimensional Minkowski space, i.e., $W_L=\{x: x^1 < -|x^0| \}$, $W_R=\{x: x^1 > |x^0| \}$. These regions are causal complements of each other. We are looking for fields $\phi$, $\phi'$ associated with $S$, such that 
\begin{equation}
   [ \phi(x), \phi'(y) ] = 0 \quad \text{if } x\in W_L, \; y \in W_R.
\end{equation}
This equation will need to be read in the sense of distributions, i.e., smeared with test functions supported in $W_L$ and $W_R$ respectively. Additionally, $\phi$ should be covariant under the proper orthochronous Poincar\'e group, and $\phi'$ should arise from $\phi$ by adjoint action of the PCT operator.

To find these fields, we follow the techniques presented in \cite{Lechner:2008,AL2017} where the construction was established for scattering functions \emph{without} poles in the physical strip. However, we shall see that certain modifications are required.

We start by specifying the underlying Hilbert space. In our case, the single-particle Hilbert space should accommodate the two species of particles, hence we consider
\begin{equation}
\mathcal{H}_1 = \bigoplus_{k = 1,2}\mathcal{H}_{1, b_k}, \quad \mathcal{H}_{1, b_k} = L^2(\mathbb{R}, d\theta).
\end{equation}
The full Hilbert space $\mathcal{H}$ is then constructed as an $S$-symmetric Fock space over $\mathcal{H}_1$. Similar to the free field, one introduces creation and annihilation operators on $\mathcal{H}$ (labeled by the particle species) which now fulfill twisted commutation relations involving the $S$-matrix:
\begin{align}
z^\dagger_{b_k}(\theta)z^\dagger_{b_\ell}(\theta') &= S^{b_\ell b_k}_{b_k b_\ell}(\theta -\theta')z^\dagger_{b_\ell} (\theta')z^\dagger_{b_k}(\theta),\\
z_{b_k}(\theta)z_{b_\ell}(\theta') &= S^{b_\ell b_k}_{b_k b_\ell}(\theta -\theta')z_{b_\ell}(\theta')z_{b_k}(\theta),\\
z_{b_k}(\theta)z^\dagger_{b_\ell}(\theta') &= S^{b_k b_\ell}_{b_\ell b_k}(\theta' -\theta)z^\dagger_{b_\ell} (\theta')z_{b_k} (\theta)+\delta_{b_k b_\ell}\delta(\theta -\theta')\pmb{1}_{\mathcal{H}}.
\end{align}
$\mathcal{H}$ is equipped with a unitary representation of the proper orthochronous Poincar\'e group which preserves particle numbers, and the antiunitary CPT operator $J$; we denote by $J_1$ its restriction to $\mathcal{H}_1$.

Now we are ready to define our quantum field: For $f \in \bigoplus _{k=1,2}\mathcal{S}(\mathbb{R}^2) $, we set
\begin{equation}
\phi(f) := z^\dagger(f^+)+ z(J_1f^-) 
       = \sum_{k=1,2} \int d\theta\, \left( f^+_{b_k}(\theta) z^\dagger_{b_k}(\theta)+ (J_1f^-)_{b_k} (\theta)z_{b_k}(\theta) \right),
\end{equation}
where $f^\pm$ indicates the Fourier transform with positive and negative frequencies, 
\begin{equation}
f^\pm_{b_k}(\theta) := \frac{1}{2\pi}\int d^2x\; e^{\pm ip_{b_k}(\theta)\cdot x} f_{b_k}(x). 
\end{equation}
Further, we define the \emph{reflected field} as $\phi'(g) := J \phi(g_j)J$, where $(g_j)_{b_k}(x) := \overline{g_{b_k}(-x)}$. 

These fields $\phi(f)$, $\phi'(g)$ are, as expected, not strictly local. The analogous fields for scalar $S$-matrices without bound states can however shown to be wedge-local \cite{Lechner:2008}. To compare this with our case, let us compute their commutator: it turns out to be
\begin{align}
&([\phi'(g), \phi(f)]\Psi_n)^{b_{k_1}, \ldots, b_{k_n}}(\theta_1,\cdots,\theta_n) \nonumber\\
&\quad=\;\int d\theta'\, \left(  g^-_{b_1}(\theta') \left(\prod_{p=1}^n S^{b_{k_p} b_1}_{b_1 b_{k_p}}(\theta' -\theta_p)\right) f^+_{b_1}(\theta')
- g^+_{b_1}(\theta') \left(\prod_{p=1}^n  \overline{S^{b_{k_p} b_1}_{b_1 b_{k_p}}(\theta' -\theta_p)}\right) f^-_{b_1}(\theta')\right) \nonumber\\
&\quad\quad\quad\quad\times(\Psi_n)^{b_{k_1}, \ldots, b_{k_n}}(\theta_1, \ldots, \theta_n).
\end{align}
One now asks whether this expression vanishes when $\operatorname{supp} f \subset W_L$ and $\operatorname{supp} g \subset W_R$. 
In fact, if $S$ were analytic in the physical strip, then  the integration contour in one of the summands could be shifted by $i\pi$ using the Cauchy theorem, and the commutator would then vanish as a consequence of crossing symmetry.
In the presence of poles of $S$ in the physical strip, however, the Cauchy theorem yields contributions from the residues of the integrand. Hence to achieve wedge commutativity, we have to modify the field with an additional term whose commutator will cancel these contributions. To that end we introduce the bound state operator $\chi(f)$, whose action on a one-particle vector (in a suitable domain, see below) is defined as follows:
\begin{equation}\label{eq:chi}
 (\chi_{1}(f)\xi)_{b_k} (\theta) := \left\{\begin{array}{ll}
                                          -i \eta^{b_1}_{b_1 b_2} f^+_{b_1} (\theta + i\theta_{(b_1 b_2)}^{b_1} ) \xi_{b_2} (\theta - i\theta_{(b_2 b_1)}^{b_1}) & \text{ if } k = 1,\\ \\
-i \eta^{b_2}_{b_1 b_1} f^+_{b_1} (\theta + i\theta_{(b_1 b_1)}^{b_2} ) \xi_{b_1} (\theta - i\theta_{(b_1b_1)}^{b_2}) & \text{ if } k = 2.
\end{array}\right.
\end{equation}
Here the matrix elements $\eta^{b_1}_{b_1 b_2},\eta^{b_2}_{b_1 b_1} $ are related to the residues of S at the poles in the strip, see \cite[Eq.~(3)]{CTsine} for the precise expressions. It is at this point that the positivity of residue condition \eqref{eq:respositive} is needed, since for the following construction to work, the $\eta$ need to be defined as proportional to $(-i \operatorname{res} S )^{1/2}$. 
The field $\chi(f)$ is then defined by lifting to second quantization, 
\begin{equation}
   \chi_n(f) := n P_n (\chi_1(f) \otimes \pmb{1} \otimes\cdots\otimes \pmb{1})P_n, \quad
 \chi(f) := \bigoplus_{n=0}^\infty \chi_n(f),
\end{equation}
where $P_n$ is the projection onto the $S$-symmetric part of the Fock space.
A corresponding ``reflected'' field is defined as $\chi'(g):=J\chi( g_j) J$.
 
The domain of the operators $\chi(f)$ and $\chi_1(f)$ is rather subtle. 
For the expression \eqref{eq:chi} to be in $L^2$, the vectors $\xi_{b_2}$, $\xi_{b_1}$ must be analytic until $\mathbb{R}- i\theta_{(b_2 b_1)}^{b_1}$ and $\mathbb{R}- i\theta_{(b_1b_1)}^{b_2}$, respectively, with their $L^2$-norm uniformly bounded in those strips.
The domain of $\chi(f)$ is even more difficult to describe; it consists of $n$-particle vectors that have such analytic continuations in each variable, but also possess certain zeros that compensate poles of $S$ in the physical strip, occurring in the operator $P_n$. (See \cite[Proof of Prop.~3.1]{CT17}.)

Now the main result of \cite{CTsine} is as follows. For the field defined as
\begin{equation}
\tilde\phi(f) = \phi(f) + \chi(f)
\end{equation}
and its reflected field $\tilde{\phi}'(g)$, we can show that those components corresponding to elementary particles (namely, of type $b_1$) commute weakly on a dense domain $D \subset \operatorname{dom}(\tilde\phi(f))\cap \operatorname{dom}(\tilde\phi'(g))$. (Note that $\operatorname{Dom}(\tilde\phi(f))= \operatorname{Dom}(\chi(f))$.)
More precisely:
\begin{theorem}
Let $f$ and $g$ be test functions supported in $W_L$ and $W_R$, respectively, and with the property that $f=f^\ast$ and $g=g^\ast$.
Furthermore, assume that $f,g$ have components $f_{b_k} =0$ and $g_{b_k} =0$ for $k \neq 1$.
Then, for each $\Phi, \Psi\in D$, we have
 \begin{equation}
 \langle \tilde\phi(f)\Phi, \tilde\phi'(g) \Psi \rangle = \langle \tilde\phi'(g) \Phi, \tilde\phi(f) \Psi \rangle.
 \end{equation}
\end{theorem}

In the proof of this theorem, the commutator $[\tilde\phi(f), \tilde\phi'(g)]$ (meant always in a weak sense) gives the contributions
$[\phi(f), \phi'(g)]$, $[\chi(f), \chi'(g)]$, $[\phi(f), \chi'(g)]$ and $[\chi(f), \phi'(g)]$. 
We can show that the last two are independently zero, while the first two cancel each other due to the chosen form of the $\chi(f)$. 

\section{Strictly local observables: Open questions}

Having constructed a wedge-local field, one would now like to proceed to strictly localized observables, either algebras of bounded operators localized in the intersection of wedges, i.e., in double cones \cite{Lechner:2008}, or pointlike fields affiliated with these algebras \cite{BostelmannCadamuro:examples}. The approach would be to define von Neumann algebras associated with $W_L$ and $W_R$, generated by bounded functions of the $\tilde \phi(f)$ and $\tilde \phi'(g)$, respectively, and corresponding algebras for shifted wedges by covariance. Then one would use abstract methods of Tomita-Takesaki theory to show that the intersection of two such algebras is sufficiently large.

For the present model, however, these questions are not settled at this point. We highlight some of the difficulties.

First, the field $\tilde\phi(f)$ is densely defined and symmetric on a suitable domain of vectors, but it is not self-adjoint on a naive domain. For defining bounded functions of the fields, one needs to show the existence of self-adjoint extensions of ̃$\tilde\phi(f)$ and ̃$\tilde\phi'(g)$, and then select those that strongly commute. This is a highly non-trivial task, but some progress has been made in the Bullough-Dodd model \cite{Tanimoto:2016}.

Second, the nontriviality of double-cone algebras was shown for certain scalar models without bound states \cite{Lechner:2008,AL2017} using the so-called ``modular nuclearity condition'' for the von Neumann algebras of wedges. Notwithstanding the construction of these algebras in our case, techniques for showing the nuclearity conditions are still being investigated.

As a final step, once local operators are fully constructed, it would be important to verify that the $S$-matrix which results by Haag-Ruelle scattering theory is indeed factorizing, and that the two-particle scattering matrix coincides with the function $S$ in \eqref{eq:sgdeformed} that our construction started from. This is another point of our ongoing work.

\section{Conclusions}

We have investigated the construction of the Sine-Gordon model \cite{BabujianFringKarowskiZapletal:1993,Babujian2002} in the operator-algebraic framework. More specifically, we have considered a deformed version of the original $S$-matrix by multiplying with a CDD factor for a certain range of the coupling constant, not only limiting the number of particle species involved but also adjusting the signs of the residues of $S$. This allowed us to establish a pair of wedge-local quantum fields, which commute at least weakly (on a suitable dense domain) at spacelike distance of the wedges if their components are restricted to elementary particles ($b_1$). For a full construction, some steps are still missing, in particular strong commutativity of the fields.

An extension to other ranges of the coupling constant, as well as the construction of the usual Sine-Gordon model without modifications by CDD factors, have not been treated in the algebraic framework yet; this remains challenging due to the subtle pole structure of the $S$-matrix and the sign of its residues as mentioned above. We hope to return to this problem in the near future.


\section*{Acknowledgements}

The author is supported by the Deutsche Forschungsgemeinschaft (DFG) within the Emmy Noether grant CA1850/1-1. 


\section*{References}

\bibliographystyle{iopart-num} 
\bibliography{integrable}

\end{document}